# Two-Carrier Transport Induced Hall Anomaly and Large Tunable Magnetoresistance in Dirac Semimetal $Cd_3As_2$ Nanoplates

Cai-Zhen Li, Jin-Guang Li, Li-Xian Wang, Liang Zhang, Jing-Min Zhang, Dapeng Yu, and Zhi-Min Liao*

State Key Laboratory for Mesoscopic Physics, School of Physics, Peking University, Beijing 100871, P.R. China

Electron Microscopy Laboratory, School of Physics, Peking University, Beijing 100871, P.R. China

Collaborative Innovation Center of Quantum Matter, Beijing 100871, P.R. China

*E-mail: liaozm@pku.edu.cn

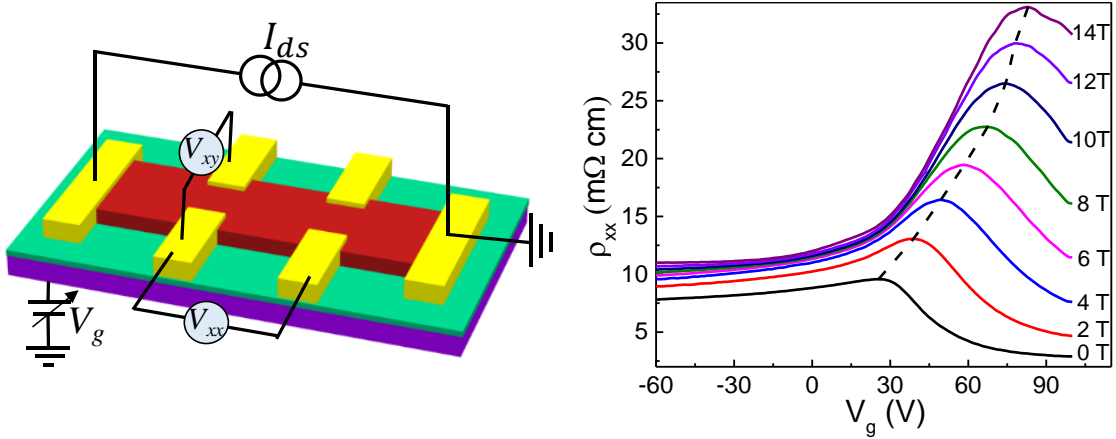

**ABSTRACT:** $Cd_3As_2$ is a model material of Dirac semimetal with a linear dispersion relation along all three directions in the momentum space. The unique band structure of $Cd_3As_2$ makes it with both Dirac and topological properties. It can be driven into a Weyl semimetal by the symmetry breaking or a topological insulator by enhancing the spin-orbit coupling. Here we report the temperature and gate voltage dependent magnetotransport properties of $Cd_3As_2$ nanoplates with Fermi level near the Dirac point. The Hall anomaly demonstrates the two-carrier transport accompanied by a transition from n-type to p-type conduction with decreasing temperature. The carrier-type transition is explained by considering the temperature dependent spin-orbit coupling. The magnetoresistance exhibits a large non-saturating value up to 2000% at high temperatures, which is ascribed to the electron-hole compensation in the system. Our results are valuable for understanding the experimental observations related to the two-carrier transport in Dirac/Weyl semimetals, such as $Na_3Bi$, $ZrTe_5$, TaAs, NbAs, and $HfTe_5$.

**KEYWORDS:** Dirac semimetal**,** magnetoresistance**,** two-band transport**,** temperature dependence**,** Hall resistance

Three-dimensional (3D) Dirac semimetal[1-3] is a quantum material, where the conduction bands and valence bands touch at discrete points, known as Dirac points. In momentum space, it has linear dispersion along all three directions near the Dirac points, which are protected by the rotational crystalline symmetry.[3,4] Upon breaking of time-reversal symmetry or spatial inversion symmetry, the Dirac point splits into a pair of Weyl nodes with opposite chiralities, thus the Dirac semimetal changes into Weyl semimetal.[5,6] The distinct electronic structures of Dirac semimetal give rise to many other topological phases, for examples, topological insulator[7,8] and topological superconductor.[9] Recently, the negative magnetoresistance (MR) has been observed in Dirac semimetal[10-13] and Weyl semimetal,[14] which is attributed to the chiral anomaly in the presence of parallel electric field and magnetic field. [15,16] In the family of 3D Dirac semimetal, $Cd_3As_2$ is a model material that has a pair of Dirac points near the $\Gamma$ point in the Brillouin zone.[3,17-19] Angle-resolved photoemission spectroscopy (ARPES) and scanning tunneling microscopy both reveal the linear dispersion near the Dirac points in $Cd_3As_2$.[17-20] Transport measurements demonstrate the ultrahigh carrier mobility in $Cd_3As_2$.[21-25] The exotic topological surface states have been studied *via* Aharonov-Bohm oscillations in thin $Cd_3As_2$ nanowires with low carrier density, making it accessible to the transport properties of the surface Fermi arcs. [26] Large linear positive MR has been observed in $Cd_3As_2$ bulk materials with high carrier concentration under perpendicular magnetic fields. [25,27] Besides, large MR has also been observed in $Cd_3As_2$ with Fermi surface close to the Dirac points.[28] Generally, the MR behaves as $(1 + \mu B^2)$ under weak magnetic fields, where $\mu$ is the carrier mobility. While in semimetals, such as high-purity graphite and bismuth, large MR under relatively low magnetic fields is ascribed to the two-carrier transport. [29] Under high magnetic fields, the large MR eventually saturates due to the breaking of the electron-hole balance. However, in $WTe_2$, the large non-saturating MR can be maintained even under strong magnetic fields, benefitting from the nearly perfect electron-hole compensation.[30] Therefore, it is highly desirable to investigate the relationship between the large MR and two-carrier transport in $Cd_3As_2$ system.

Here we report the magnetotransport properties of $Cd_3As_2$ nanoplates. The advantage of $Cd_3As_2$ nanoplates compared with bulk crystals is that the carrier density can be modulated by gate voltage to realize ambipolar transport. Thus, it is more convenient to investigate the two-carrier transport behavior. The MR exhibits a saturating value ~50% at low temperatures and large non-saturating value up to 2000% at high temperatures in $Cd_3As_2$ nanoplates. By employing gate voltage modulation and Hall measurements, large asymmetry of electron and hole transport has been revealed. By applying positive gate voltage, the MR can be tuned from a small saturating value to a large non-saturating value at low temperatures. Besides, due to the asymmetric mobilities of electrons and holes, the maximum of longitudinal resistivity in the transfer curves shifts towards to positive gate voltage with increasing magnetic field. The results reveal that the electron-hole compensation may be responsible for the large non-saturating MR.

**Results**

    **Nanoplate Characterizations.** The as-grown $Cd_3As_2$ nanoplates are single crystalline.[31] As shown in **Figure 1a**, the scanning electron microscopy (SEM) image indicates that the lateral dimension of the nanoplates ranges from several micrometers to tens of micrometers. A corner of a typical nanoplate is shown by the transmission electron microscopy (TEM) image in **Figure 1b**. The high-resolution TEM image in **Figure 1c** shows an interplanar spacing of ~0.23 nm, indicating the ($1\overline{1}0$) edge direction of the nanoplate. The selected area electron diffraction (SAED) pattern is shown in **Figure 1d**. By calculating the distance from each diffraction point to center point and their relative orientations, we determined the ($4\overline{4}0$), ($0\overline{4}8$), ($\overline{4}08$) crystal planes, as marked in **Figure 1d**. According to the crystallography calculation, the (112) top surface plane of tetragonal $Cd_3As_2$ projection with [221] zone axis can be determined. The energy-dispersive X-ray spectroscopy (EDS) of the nanoplate is shown in **Figure 1e**. The quasi-quantitative analysis of the EDS indicates that the Cd to As atomic ratio is 58.3: 41.7 with an uncertainty less than 1%, consistent with the

chemical composition of $Cd_3As_2$. The thickness of the synthesized nanoplates ranges from 200 nm to 700 nm roughly, as shown by the atomic force microscopy image in Figure S1. As $Cd_3As_2$ is not a layered material, it is difficult to get thinner nanoplates down to a few unit cell layers in the chemical vapor deposition (CVD) growth system.

**Temperature Dependent Electron-Hole Transport. Figure 2a** shows the schematic diagram of the nanoplate device with six-terminal Hall-bar geometry. The Si substrate with 285 nm $SiO_2$ thin layer serves as the back gate. **Figure 2b** shows the temperature dependence of the longitudinal resistivity, which exhibits a semiconducting behavior at high temperatures and a metallic behavior at low temperatures. This $\rho_{xx} - T$ dependence is distinct with that of $Cd_3As_2$ bulk crystals. The semiconducting behavior of $Cd_3As_2$ is ascribed to the low carrier density and can be interpreted by the thermal activation mechanism. The activation energy $E_a$ is extracted to be 19.53 meV according to the Arrhenius plot at high temperatures (inset in **Figure 2b**). Similarly, the semiconducting $\rho_{xx} - T$ curves have also been observed in Sb doped $Bi_2Se_3$ topological insulator with a low carrier density. [32]

Figure 3a shows the Hall resistivity under small magnetic fields at $V_g = 0$ V and at 1.5 K. The nonlinear Hall curve is repeatable for the sweeping cycles without distinguishable hysteresis, and it is symmetric under negative and positive magnetic field. The evolution of the Hall curves with temperature (**Figure 3b**) clearly shows a transition from p-type to n-type conduction with increasing temperature. The observed Hall anomaly is a characteristic of two-carrier transport, which can be described by the two-carrier model: [33-35]

$$\rho_{xy} = \frac{1}{e} \frac{(n_h \mu_h^2 - n_e \mu_e^2) + \mu_h^2 \mu_e^2 B^2 (n_h - n_e)}{(n_h \mu_h + n_e \mu_e)^2 + \mu_h^2 \mu_e^2 B^2 (n_h - n_e)^2} B, \qquad (1)$$

$$\rho_{xx} = \frac{1}{e} \frac{(n_h \mu_h + n_e \mu_e) + (n_e \mu_e \mu_h^2 + n_h \mu_h \mu_e^2) B^2}{(n_h \mu_h + n_e \mu_e)^2 + \mu_h^2 \mu_e^2 B^2 (n_h - n_e)^2}, \qquad (2)$$

where $n_e(n_h)$ and $\mu_e(\mu_h)$ are the carrier density and mobility of electrons (holes), respectively. According to the Eq. (1), the Hall resistivity reverses its sign at a critical magnetic field $B_C = \sqrt{\frac{n_e \mu_e^2 - n_h \mu_h^2}{\mu_h^2 \mu_e^2 (n_h - n_e)}}$. In case of $\mu_e > \mu_h$, the increase of electron concentration with increasing temperature will lead to the increase of $B_C$, which is consistent with our results in **Figure 3b**. The magnetic field dependence of the Hall resistivity can be well fitted according to Eq. (1) at representative 80 K and 100 K, as shown in **Figure 3c**.

**Figures 3d-e** show the magnetic field dependence of MR (defined as $\frac{\rho_{xx}(B) - \rho_{xx}(0)}{\rho_{xx}(0)} \times 100\%$) at various temperatures from 1.5 K to 300 K. At low temperatures, the MR first increases slightly with increasing magnetic field and then tends to saturation. At 14 T, the MR increases from 43% at 1.5 K to 914% at 150 K. While at 200 K and 300 K, the MR shows a quadratic behavior without any tendency to saturation. Largest MR up to 2000% is observed at 200 K, which is 50 times larger than that at 1.5 K. The observed large non-saturating MR at high temperatures may be due to the thermally activated electrons, leading to the two-carrier transport. According to Eq. (2), the two-carrier transport results in a non-saturating quadratic MR with $\frac{\Delta \rho_{xx}}{\rho_0} = \frac{n_h \mu_h n_e \mu_e (\mu_h + \mu_e)^2}{(n_h \mu_h + n_e \mu_e)^2} B^2$, which is consistent with our results at high temperatures.

To reveal the two-carrier transport in the $Cd_3As_2$ nanoplate, the Kohler's plots are presented in **Figure 3f**. The MR at different temperatures could be rescaled by the

Kohler's plot:[33,36]

$$\frac{\Delta R_{xx}(B)}{R_{xx}(0)} = F(\frac{B}{R_{xx}(0)}). \quad (3)$$

For the classical $B^2$ dependence of MR, $\Delta R_{xx}(B)/R_{xx}(0) \propto [B/R_{xx}(0)]^2$. In case of single type of carrier dominant transport, the Kohler's plots at different temperatures should overlap with each other. As shown in **Figure 3f**, at low temperatures the curves are collapsed together, indicating the hole-dominant transport; at medial temperatures the curves separate from each other due to the gradual increase of electron concentration with increasing temperature; and at high temperatures the curves re-collapse to a single curve, suggesting the electron-dominant transport. Despite the coexistence of electron and hole at high temperatures, the electron-dominant transport is mainly due to the much higher mobility of electron than that of hole.

**Gate Tunable Electron-Hole Transport. Figure 4a** shows the longitudinal conductivity as a function of gate voltage at 1.5 K without external magnetic fields. A minimum of longitudinal conductivity (Dirac point $V_D$) at $V_g \sim 5$ V was observed at 1.5 K, indicating the intrinsic p-type doping at low temperatures. The conductivity increases sharply with increasing the positive gate voltage. For the part of negative gate voltage, the conductivity changes slightly. The large asymmetry transport is due to the different mobilities of electrons and holes. The carrier mobility can be estimated by linear fitting of $\sigma_{xx}$ through $\mu = \frac{\partial \sigma_{xx}}{\partial V_g} \frac{dt}{\varepsilon_0 \varepsilon_{ox}}$, where $d$ and $t$ are the thickness of nanoplate and oxide layer, respectively, $\varepsilon_{ox} = 3.9$ is the dielectric constant of $SiO_2$. According to $\sigma = ne\mu$, the carrier density is roughly estimated and presented in **Figure 4b**. At $V_g = 0$ V, the nanoplate is a p-type system. By applying positive gate voltage, there will be an n-type conduction layer near the $Cd_3As_2/SiO_2$ interface (**Figure S2**). However, as the nanoplate is relatively thick (~350 nm), the gate electrical field can not penetrate through the sample due to the screening effect. Therefore, at positive gate voltages, it is a two-channel system with an inversion layer near the $Cd_3As_2/SiO_2$ interface.[37,38] Two parallel conduction path model gives similar form of formula with that in Eqs.(1)-(2), with subscript parameters corresponding to the conduction layer.[37]

The longitudinal conductivity as a function of gate voltage at all temperatures is shown in **Figure 4c**. The Dirac point $V_D$ shifts towards negative gate voltage with increasing temperature, suggesting the thermal activation of electrons. Further increasing temperature up to 100 K, the Dirac point can not be found clearly in the $V_g$ ranging from -60 V to 60 V. Meanwhile, the conductivity increases monotonously with increasing gate voltage (see **Figure S3**), demonstrating a clear n-type feature.[39] According to the gate modulation formula, the electron mobility $\mu_e = 1.3 \times 10^4$ cm$^2$/Vs is estimated from the transfer curve at 1.5 K. The hole mobility is estimated to be $\mu_h = 9 \times 10^2$ cm$^2$/Vs. The low hole mobility may result from the severe scatterings and the low Fermi velocity in the valence band.[18,19] The carrier mobility is further decreased with increasing temperature, as shown in **Figure 4d**.

The gate-tunable MR behaviors were carefully investigated with magnetic field perpendicular to the substrate. **Figure 5a** shows the MR behaviors under various gate voltages at 1.5 K. One can clearly see that the MR increases tardily under low magnetic fields and then gradually saturates with ~ 42.5% under 14 T at $V_g$ = -60 V,

-40 V, -20 V and 0 V. The MR is enhanced by applying positive gate voltage and up to ~200% at $V_g$ = 60 V. To reveal the mechanism of the MR improvement, the corresponding Hall resistivities were measured and presented in **Figure 5b**. The evolution of nonlinear Hall resistivity with gate voltage (**Figure S4**) clearly demonstrates the injection of electrons by positive gate voltage. The high electron mobility gives rise to the negative slopes of $\rho_{xy} \sim B$ curves under low magnetic fields. While the positive slopes of the $\rho_{xy} \sim B$ curves under high magnetic fields indicate the hole-dominant $\rho_{xy}$. The critical field $B_C$ increases with increasing electron density by applying positive gate voltage, as shown in **Figure 5b**. In the presence of two parallel conduction paths, the Eq. (1) can also be used to analyze the Hall data because the tow-channel transport can be described in the frame of two-carrier model.[35,37] At high magnetic fields, the Hall coefficient approximately equals to $\frac{1}{e} \frac{1}{n_h - n_e}$. Therefore, the net carrier density $n_h - n_e$ can be extracted as a function of gate voltage, as shown in **Figure 5c**.

The MR and $\rho_{xy}$ at a fixed $V_g$ of 60 V were further measured at different temperatures (**Figure S5**). To further investigate the gate voltage modulation on MR, the MRs at $V_g$ = 0 V and 60 V under 14 T from 1.5 K to 300 K are presented in **Figure 6a**. The MR is much improved by applying 60 V gate voltage to increase the electron concentration at temperatures lower than 150 K, while slightly decreases at 200 K and 300 K. In order to reveal the origin of the distinct MR of $Cd_3As_2$ nanoplates, the temperature-dependent electron density at $V_g$ = 60 V have been calculated and presented in **Figure 6b.** The electron density increases drastically above 150 K, in consistent with the $\rho$-$T$ behavior. The exponential fitting of the $n_e$-$T$ curve gives an activation energy of 23.4 meV, which is comparable to that extracted from the $\rho$-$T$ Arrhenius plot. The density of thermally activated electron at 200 K is comparable to the density of the intrinsically doped hole at 1.5 K. Therefore, a nearly electron-hole compensation state gives rise to the largest MR at 200 K. We would like to stress here that electron-dominant transport is not in conflict with the nearly equal electron-hole populations at 200 K, considering the mobility of electron is much higher than that of hole. As further increasing temperature to 300 K, more electrons break the electron-hole compensation, leading to the attenuated MR. In addition, the decrease of MR at $V_g = 60$ V and at 200 K in **Figure 6a** is also consistent with the broken electron-hole compensation induced by the gate voltage.

**Magnetic Field Modulated Gate Voltage Dependence.** The transfer curves of another similar device with gate voltage up to 100 V at 1.5 K are shown in **Figure 7a**. An obvious feature is that the Dirac point shifts towards to positive gate voltages with the increase of magnetic field. **Figure 7b** presents the corresponding Hall resistivity as a function of $V_g$. As sweeping $V_g$, the $\rho_{xy}$ crosses to zero and changes its sign at a critical gate voltage ($V_{th}$), indicating the change of carrier type of the dominant transport. For comparation, the $\rho_{xx}$ and $\rho_{xy}$ $vs.$ $V_g$ at the same magnetic field (2 T) are plotted together in **Figure 7c**. It shows that $\rho_{xy}$ reverses its sign at a critical $V_g$ when $\rho_{xx}$ approaches the maximum value. The $V_D$ in $\rho_{xx}$ and $V_{th}$ in $\rho_{xy}$ as a function of magnetic field are shown in **Figure 7d**. Apparently, the two series of critical points consist with each other, providing evidence for the correlation between $V_D$ shift and electron-hole compensation. The maximum of $\rho_{xx}$ is a comprehensive result of contributions of the two-carrier densities and mobilities, which is exceptionally evident in electron-hole asymmetric systems. In other words, the large MR is obtained in an electron-hole compensating state, where the Hall resistivity equals to zero. Thus, the magnetic field induced shift of $V_D$ is due to the highly asymmetric transport of electron and hole in $Cd_3As_2$ nanoplates.

The MR (14 T) = R (14 T)/R (0 T) - 1 as a function of $V_g$ is shown in **Figure 7e**. The MR dramatically increases with increasing $V_g$ after a critical point ~ 30 V, and then saturates after $V_g$ ~ 80 V. The magnetic field dependences of *MR* at different gate voltages are shown in **Figure 7f**. To make the curves under $V_g$ = -60 V and 0 V distinguishable, the magnification of the MR < 45% is shown in the inset in **Figure 7f**. At $V_g$ = 100 V, a quadratic large non-saturating *MR ~ B* dependence is observed, and the MR is up to 1000% under 14 T at 1.5 K. At $V_g$ = 100 V, the $\rho_{xy}$~*B* curve (**Figure S6**) also indicates the electron-dominant transport, and the nearly saturated $\rho_{xy}$ at around 14 T suggests the coexistence of electron and hole. **Figure 7f** shows the MR at $V_g$ = 100 V is lower than that at $V_g$ = 50 V within the magnetic field between 1 T and 5 T, which is due to the system at $V_g$ = 50 V is much closer to the Dirac point (**Figure 7d**).

**Discussion**

The semiconducting R-T behavior and carrier-type transition with temperature are commonly observed in semimetals, such as Dirac semimetal $Na_3Bi$,[10] Weyl semimetals $TaAs$,[40] $NbAs$,[41] and transition metal pentatellurides like $ZrTe_5$.[42] To be specific, it was reported that the temperature induced electronic structure change in $ZrTe_5$ can lead to the transition of conductive carrier type with decreasing temperature from 300 K to 2 K.[42] For Dirac semimetal $Cd_3As_2$, the crossed linear energy bands are resulted from the spin-orbit coupling (SOC) induced band inversion. The calculated band structures of $Cd_3As_2$ can be well fitted by eight-band model $H_8(\vec{k}) = I \otimes H_4(\vec{k}) + H_{so}$ with the SOC parameter $\Delta \sim 0.16$ eV.[3] The electronic band structures of the Dirac semimetal described by the effective eight-band model should be very sensitive to the temperature dependent spin-orbit coupling. The temperature-sensitive SOC has been reported in SrIrO3 film,[43] GaAs/AlGaAs[44] and InSb/InAlSb[45] quantum wells, and Bi films.[46] The usually neglected higher-order terms in the Rashba Hamiltonian may be responsible for the temperature-dependent Rashba coefficient $\alpha = g(1-g)\frac{\pi e h^2 \varepsilon}{4 m_e^2 c^2}$ , through the temperature-dependent Landé factor $g = g_0 + \beta T$.[43-46] It is found that the Rashba SOC coefficient $\alpha$ increases and the g factor decreases with increasing temperature.[43] Previous investigations show that the g factor in $Cd_3As_2$ depends largely on electron concentration[47,48] or energy.[48] Here, the obvious temperature-dependent carrier concentration in our $Cd_3As_2$ system may suggest the temperature dependent SOC. The variation of the spin-orbit coupling with temperature may result in the bandgap opening at high temperatures in $Cd_3As_2$, which provides a possible scenario for understanding the abnormal transport phenomena. Although the mechanism of the temperature induced carrier type transition is still not clear, the coexistence of electrons and holes in the system with low carrier density seems to be a common feature.

**Conclusions**

In summary, we have investigated the temperature dependent and gate tunable transport properties in $Cd_3As_2$ nanoplates by the magnetoresistance and Hall resistance measurements. Taking advantage of low carrier density in our samples, the ambipolar field effect is realized. The transfer curves clearly demonstrate the asymmetric mobilities of electrons and holes, consistent with the band structure observed in ARPES measurements. The Dirac point in the longitudinal resistivity corresponds to the zero Hall resistivity as revealed by the transfer curves under different magnetic fields. Large non-saturating MR is obtained in the case of balanced population of electron and hole. Our results provide a crucial step towards understanding the properties of $Cd_3As_2$ with Fermi surfaces close to the Dirac point. The further reduction of the nanoplate thickness would make the gate modulations more effective and pave the way for the realization of quantum spin Hall insulator.

Note: During the proof of this paper, we noticed several works on Hall anomaly in $HfTe_5$,[49] and the electron-hole transport in Weyl semimetals $TaAs_2$,[50] $NbAs_2$.[51]

**Methods**

**Sample Synthesis.** $Cd_3As_2$ nanoplates were synthesized by CVD method in a tube furnace. The $Cd_3As_2$ powders (Alfa Aesar 99.99% purity) were placed at the center of the furnace and silicon substrates were placed downstream of the source to collect the products. First, the furnace was flushed by Argon gas several times to get rid of the oxygen. Then the furnace was heated from room temperature to 720 ℃ in 25 minutes and kept for 10 minutes for growth. During growth, Argon carrier gas flowed at a rate of 20 sccm. Finally, the furnace was cooled down to room temperature naturally.

**Characterizations.** The synthesized nanoplates were systematically characterized *via* scanning electron microscopy (SEM FEI Nano430) and transmission electron microscopy (TEM FEI Tecnai F20), as well as energy-dispersive X-ray spectroscopy performed on TEM system.

**Device Fabrication and Measurements.** For electrical transport measurements, individual nanoplates were transferred onto silicon substrate with an oxide layer of 285 nm, which can be used as dielectric layer for back-gate modulation. Ohmic contacts were fabricated by a series of processes including e-beam lithography, electrode deposition and lift off. Transport measurements were performed in an Oxford cryostat system with magnetic field up to 14 T and temperature down to 1.5 K. The electrical signals were measured by lock-in amplifiers (Stanford SR830) at low frequency of 17.7 Hz. A Keithley 2400 source meter was used to apply gate voltage on the silicon substrate.

*Conflict of Interest:* The authors declare no competing financial interests.

*Supporting Information Available:* The Supporting Information is available free of charge on the ACS Publications website at DOI: 10.1021/acsnano.****.

*Acknowledgement.* This work was supported by MOST (Nos. 2013CB934600, 2013CB932602) and NSFC (Nos. 11274014, 11234001, 11327902).

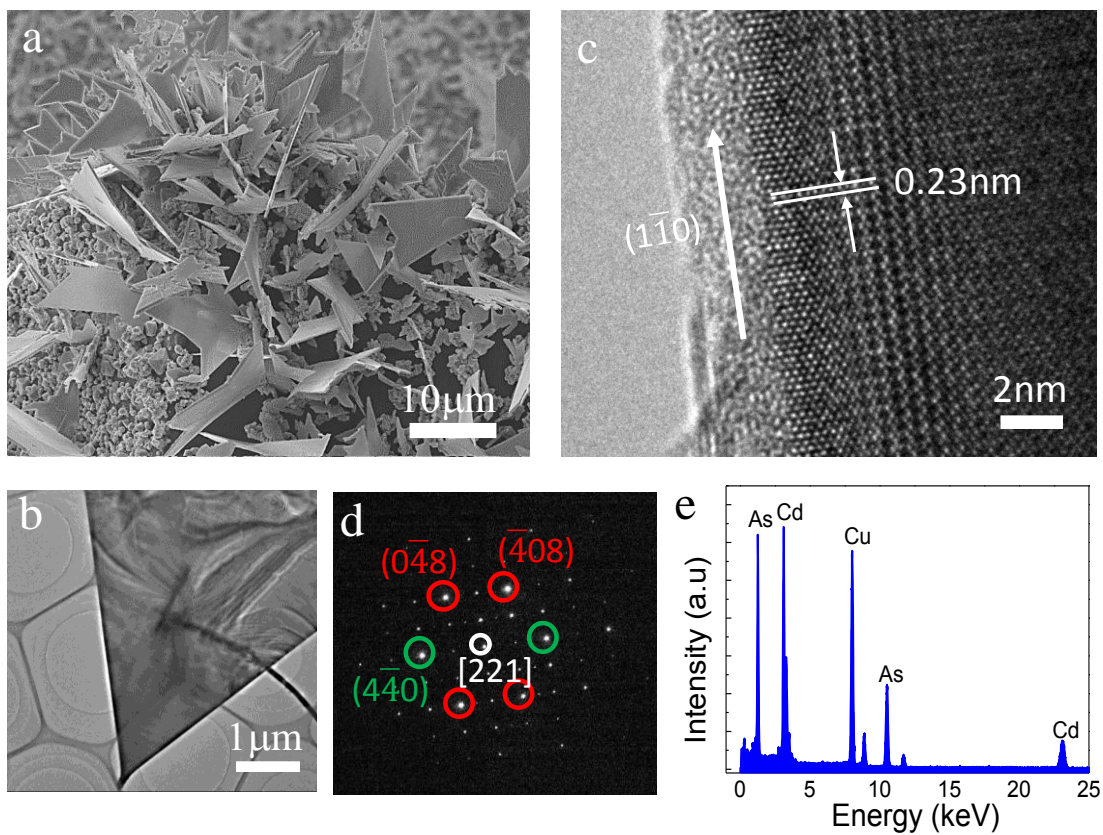

**Figure 1.** Characterization of the synthesized $Cd_3As_2$ nanoplates. (a) SEM image of the nanoplates with lateral dimensions ranging from several micrometers up to tens of micrometers. (b) The TEM image and (c) the high-resolution TEM image of a typical nanoplate. The 0.23 nm interplanar spacing indicates the ($1\bar{1}0$) direction of the nanoplate edge. (d) The SAED pattern clearly shows hexagonal symmetry of the (112) surface plane projection with [221] zone axis. (e) The EDS spectrum of the nanoplate. The quasi-quantitative analysis gives that the Cd to As atomic ratio is 58.3: 41.7 with an uncertainty less than 1%.

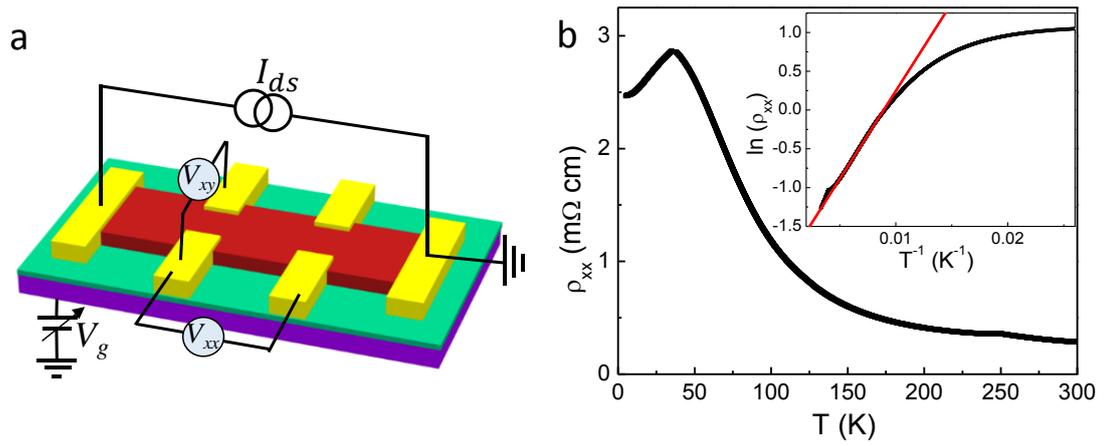

**Figure 2.** (a) Schematic diagram of the Cd$_3$As$_2$ nanoplate Hall device with a back gate terminal. (b) Temperature dependent longitudinal resistivity $\rho_{xx}$ of Cd$_3$As$_2$ nanoplate. The activation energy of 19.53 meV is deduced from the Arrhenius plot of $\rho_{xx}$ in the inset.

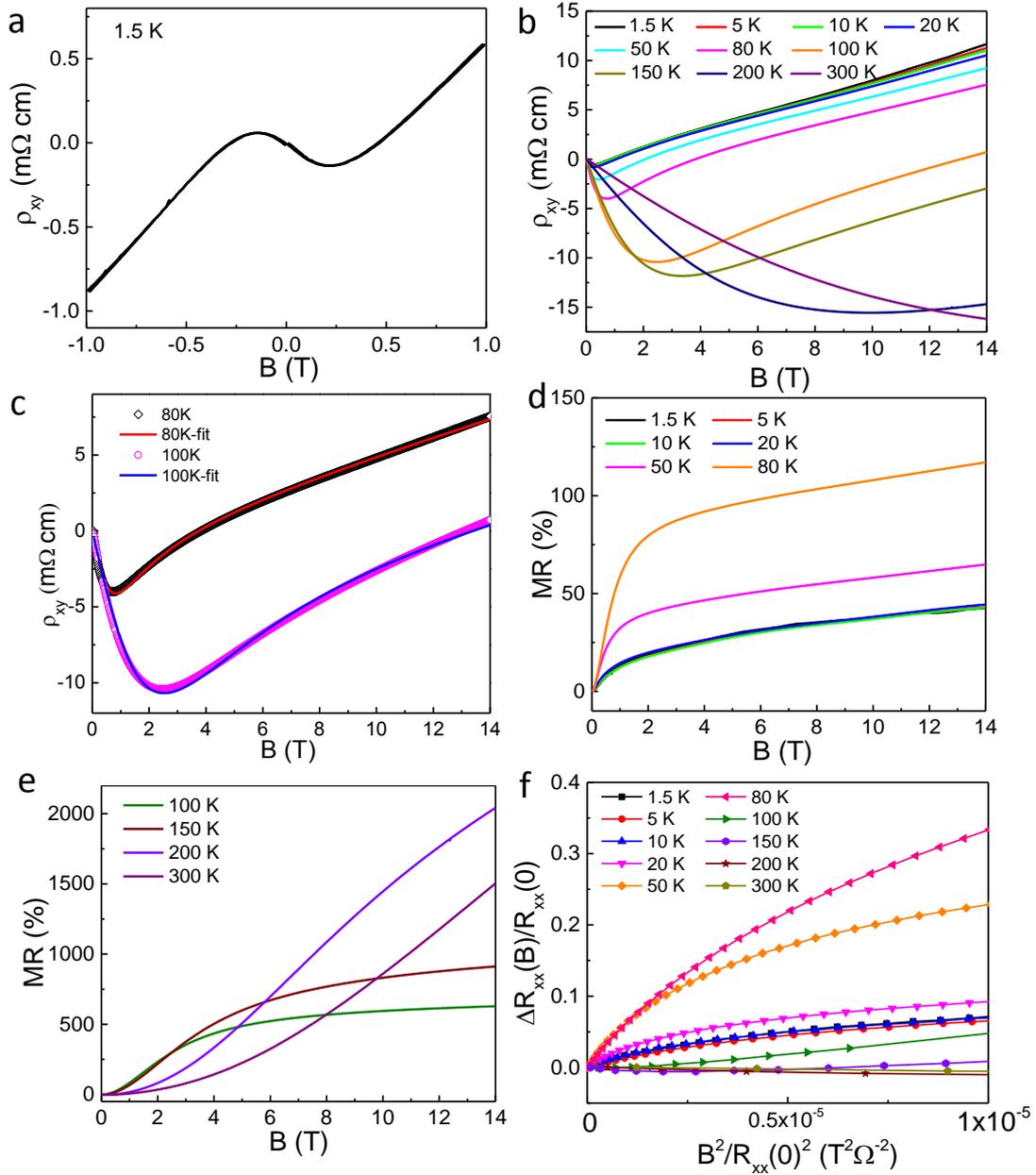

**Figure 3.** The temperature dependent magnetotransport at $V_g = 0$ V. (a) The Hall resistivity under small magnetic fields at 1.5 K. (b) The Hall resistivity at different temperatures from 1.5 K to 300 K. (c) The nonlinear fits of the Hall resistivity based on two-carrier model at representative temperatures of 80 K and 100 K. (d, e) MR measured at different temperatures from 1.5 K to 300 K. (f) The Kohler's plot of the MR curves.

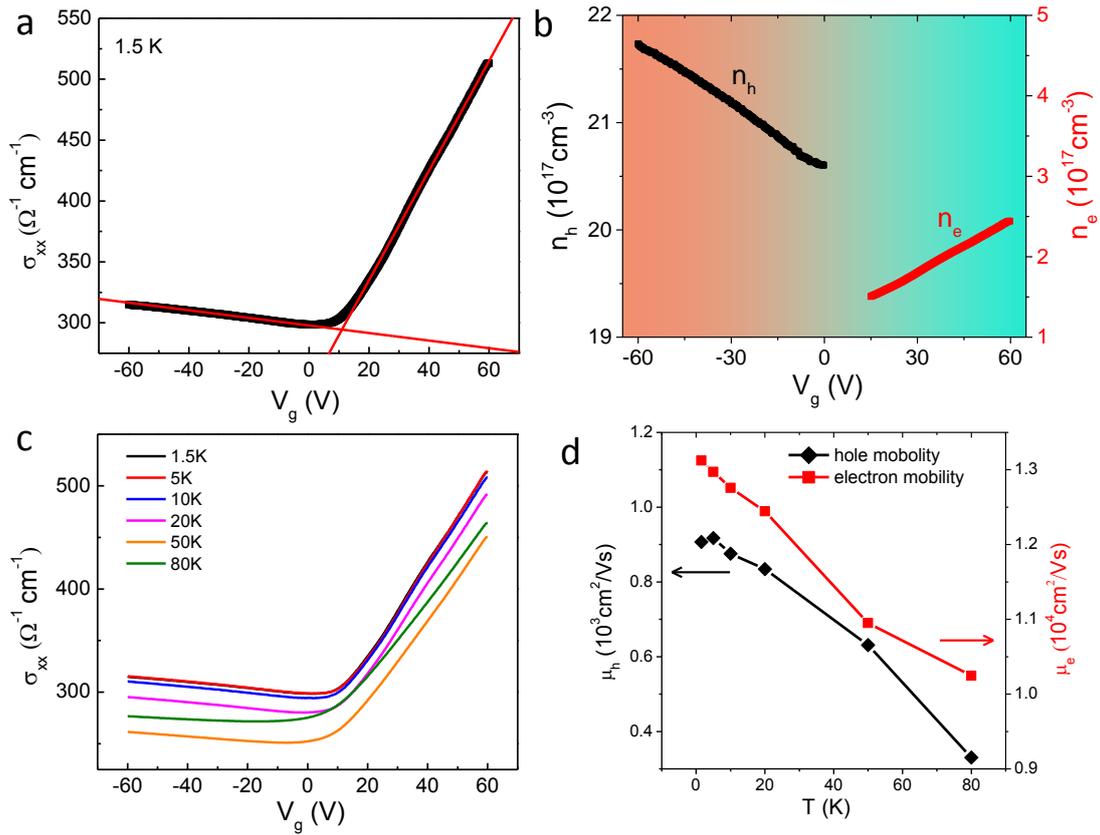

**Figure 4.** (a) The longitudinal conductivity as a function of gate voltage at 1.5 K. The $\sigma_{xx}$ reaches a minimum at around $V_g = 5$ V. (b) The hole and electron concentrations obtained from the transfer curve at 1.5 K. (c) The longitudinal conductivity as a function of gate voltage at temperatures from 1.5 K to 80 K. (d) The temperature dependence of electron and hole mobilities from 1.5 K to 80 K. The electron mobility is much larger than that of hole.

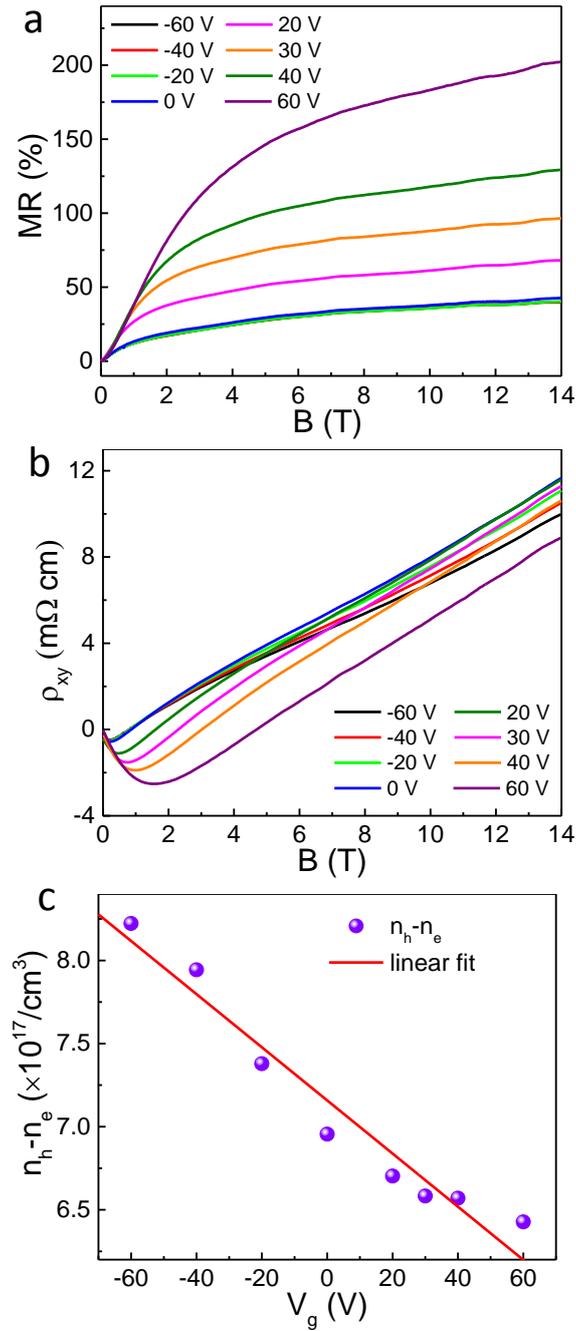

**Figure 5.** (a) The MRs under various gate voltages and (b) the corresponding Hall resistivity at 1.5 K. (c) The net carrier density as a function of gate voltage at 1.5 K of obtained from high field Hall resistivity.

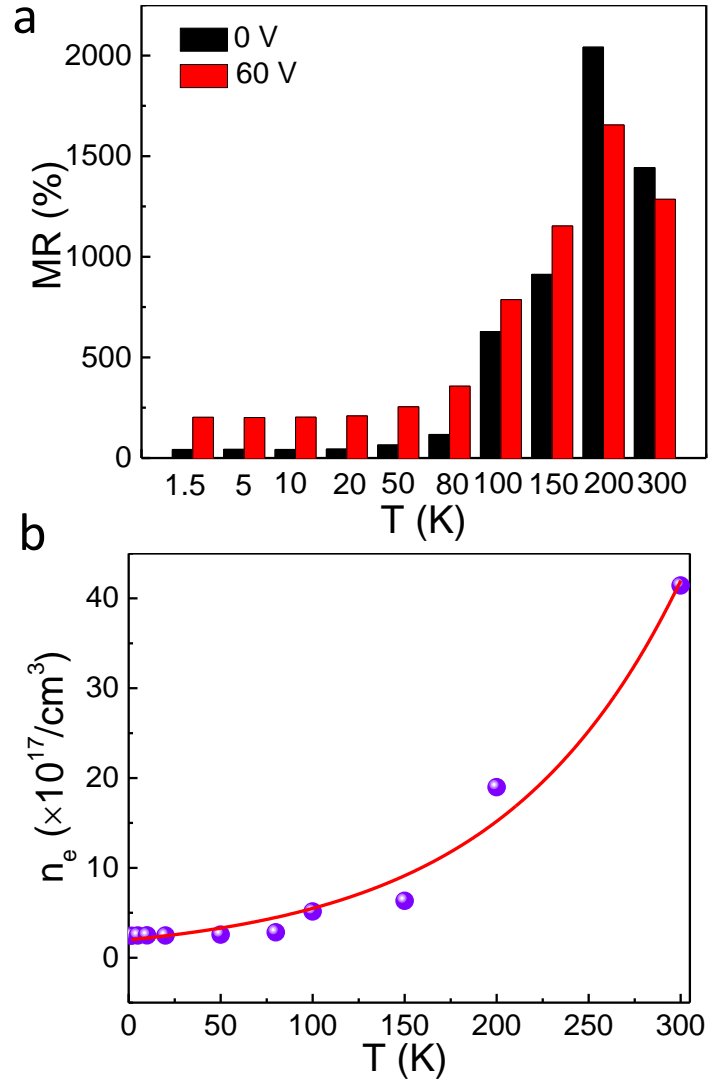

**Figure 6.** (a) The MR obtained under B = 14 T at V$_g$ of 0 V and 60 V at different temperatures from 1.5 K to 300 K. At temperatures lower than 150 K, the MR is much improved by applying 60 V gate voltage. While at 200 K and 300 K, the MR slightly decreases at V$_g$ = 60 V. (b) The electron density obtained at V$_g$ = 60 V as a function of temperature, the exponential fit of the $n_e$ - $T$ curve gives an activation energy of 23.4 meV.

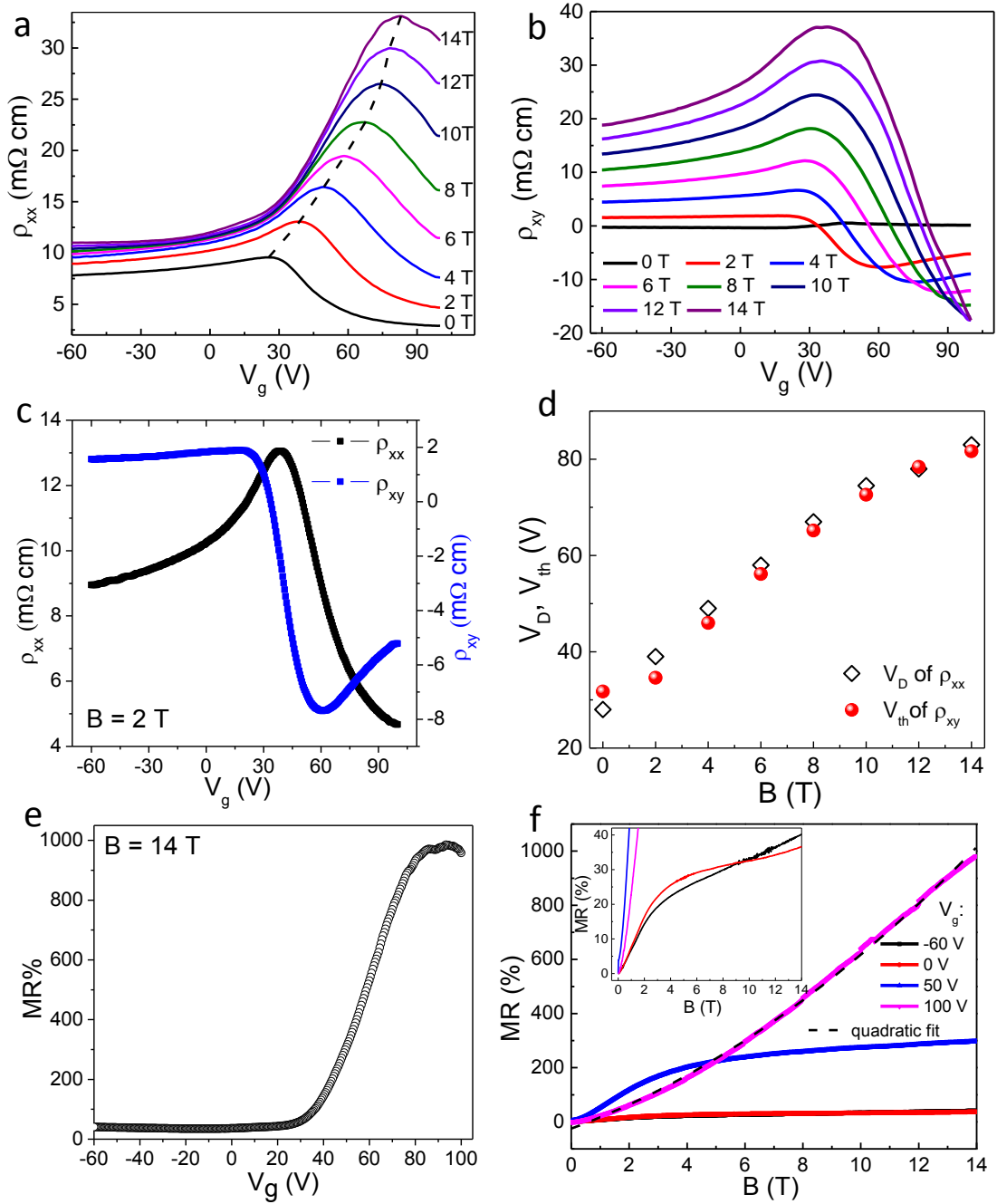

**Figure 7.** Magnetotransport of another device at 1.5 K and with gate voltage up to 100 V. (a) Transfer curves under different magnetic fields. The peak of $\rho_{xx}$, that is the Dirac point $V_D$, shifts towards to positive gate voltage with the increase of magnetic field, and (b) the corresponding Hall resistivity as a function of gate voltage under various magnetic fields. The Hall resistivity crosses to zero at a critical gate voltage $V_{th}$ for a given magnetic field. (c) The comparison of the gate voltage dependence of $\rho_{xx}$ and $\rho_{xy}$ under B = 2 T. The $\rho_{xx}$ reaches a maximum as the Hall resistivity crosses to zero. (d) The Dirac point $V_D$ of $\rho_{xx}$ and the $V_{th}$ of $\rho_{xy}$ as a function of magnetic field. The $V_D$ and $V_{th}$ have similar magnetic field dependence. (e) The MR (14 T) = R (14 T)/R (0 T) - 1 as a function of $V_g$. (f) The MR as a function of magnetic field at different gate voltages. At $V_g$ = 100 V, there is an non-saturating MR up to 1000% at 1.5 K, Inset: Magnification of MR to show clearly the curves at $V_g$ = -60 V and 0 V.

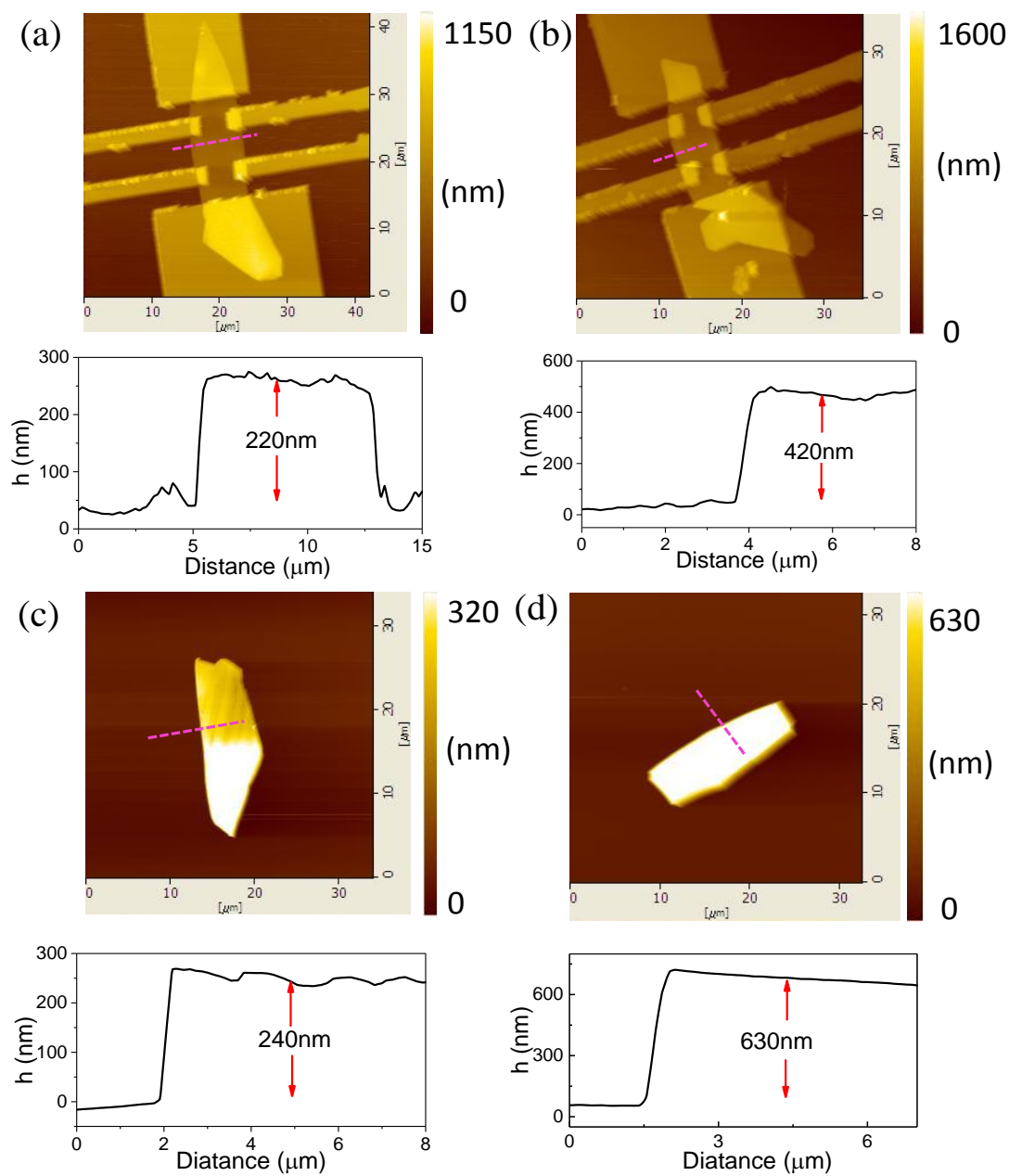

**Figure S1.** The atomic force microscopy (AFM) images and the height profiles of the Cd$_3$As$_2$ nanoplates with thickness (a) 220 nm, (b) 420 nm, (c) 240 nm, and (d) 630 nm, respectively. The thickness of the synthesized Cd$_3$As$_2$ nanoplates is in the range of 200 ~ 700 nm approximately.

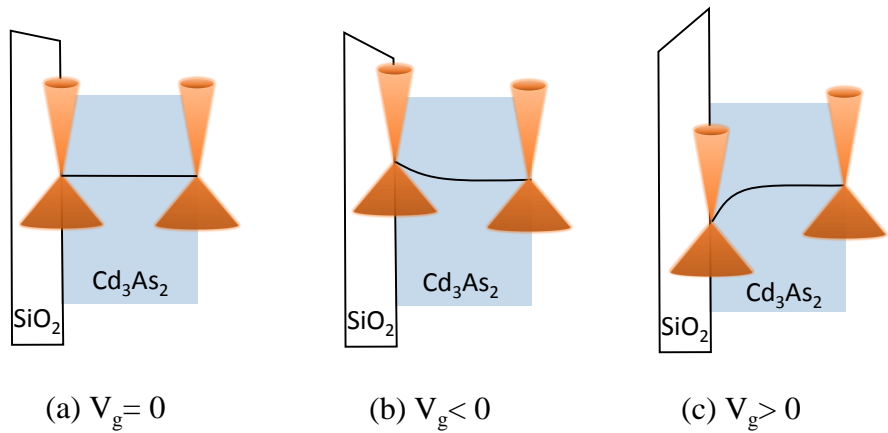

(a) $V_g = 0$       (b) $V_g < 0$       (c) $V_g > 0$

**Figure S2.** Band diagram of the quasi metal-$SiO_2$-$Cd_3As_2$ structure. (a) Without gate voltage. (b) At a negative gate voltage, a p-type accumulation layer at the bottom interface. (c) At a positive gate voltage, an n-type inversion layer at the interface. Because the nanoplate is much thicker than the width of the inversion region, the charge transport is indeed two-channel system at positive gate voltages due to the screening effect.

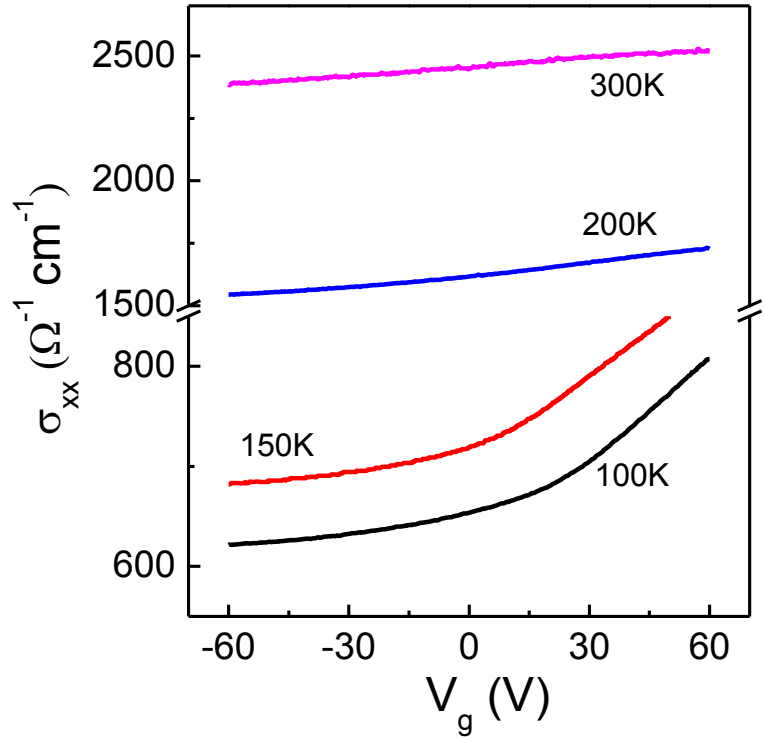

**Figure S3.** The longitudinal conductivity at different temperatures from 100 K to 300 K. The conductivity increases monotonously with increasing the gate voltage, demonstrating a clear n-type feature.

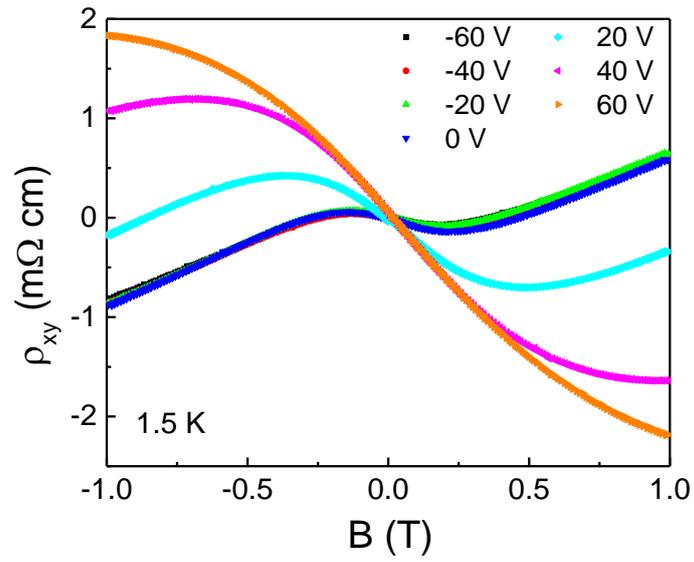

**Figure S4.** The nonlinear Hall resistivity at various gate voltages under small magnetic fields.

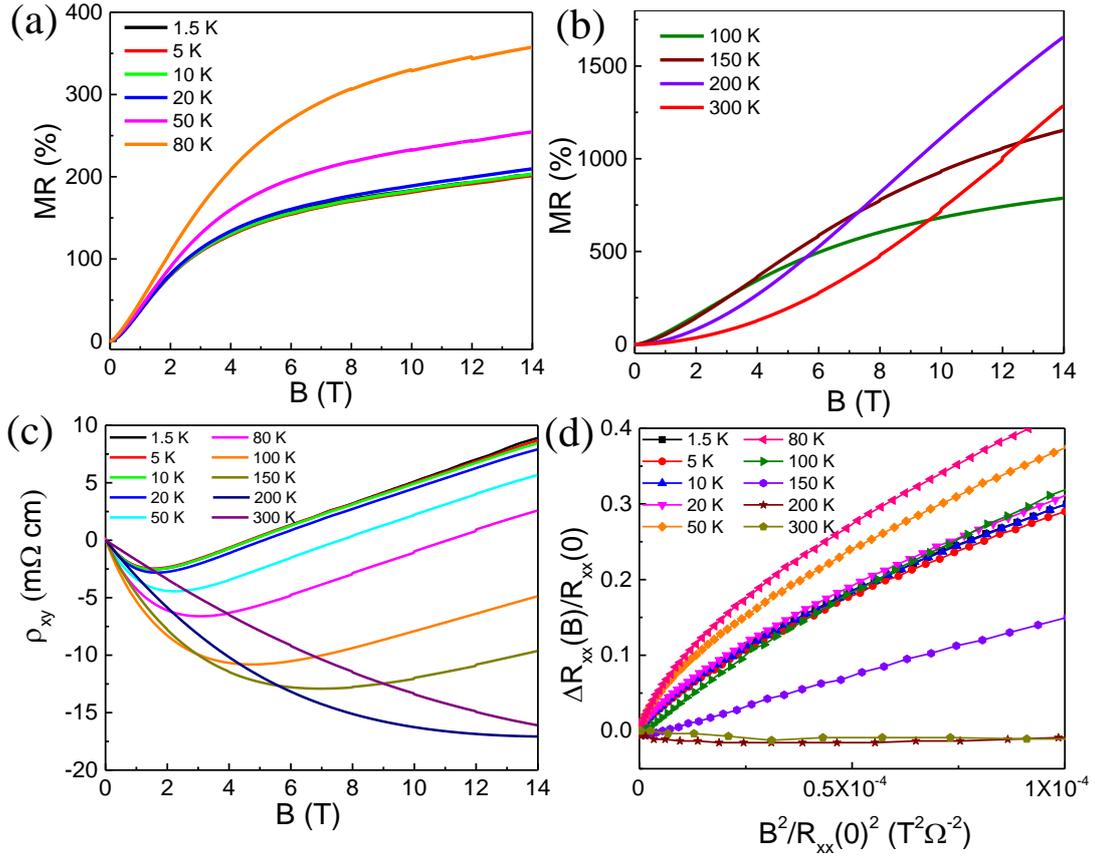

**Figure S5.** (a,b) MR measured at $V_g = 60$ V and at different temperatures from 1.5 K to 300 K and (c) the corresponding Hall resistivity. (d) The Kohler's plots of the MR curves at $V_g = 60$ V.

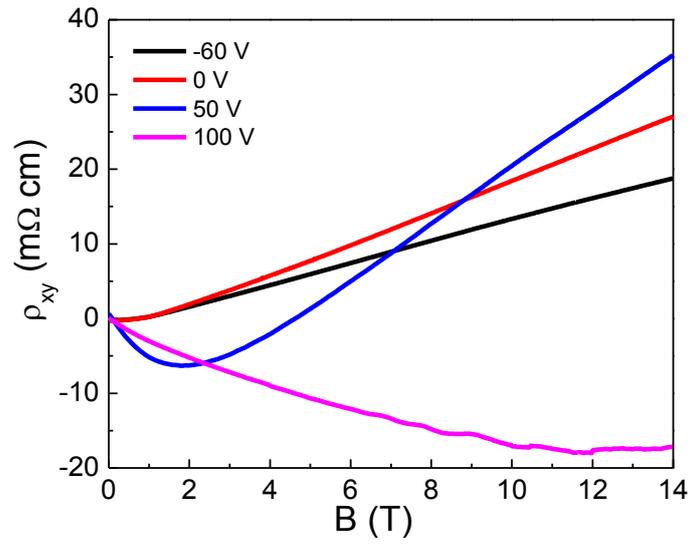

**Figure S6.** The magnetic field dependence of Hall resistivity at 1.5 K and at different gate voltages of $V_g$ = -60 V, 0 V, 50 V, and 100 V, respectively.